\documentstyle[12pt,epsf]{article}

\input epsf
\newcommand{\beq}{\begin{equation}}
\newcommand{\eeq}{\end{equation}}
\newcommand{\ba}{\begin{array}}
\newcommand{\ea}{\end{array}}
\newcommand{\bea}{\begin{eqnarray}}
\newcommand{\eea}{\end{eqnarray}}
\newcommand{\bean}{\begin{eqnarray*}}
\newcommand{\eean}{\end{eqnarray*}}
\newcommand{\nin}{\noindent}
\newcommand{\ra}{\rightarrow}
\newcommand{\ee}{{\rm e}}

\newcommand{\xx}{{\bf x}}
\newcommand{\XX}{{\bf X}}
\newcommand{\tXX}{\tilde{\XX}}
\newcommand{\cXX}{\check{\XX}}
\newcommand{\eps}{\varepsilon}

\newcommand{\const}{{\rm const.}}
\makeatletter \ifcase\@ptsize \font\teneufm=eufm10
\font\seveneufm=eufm7 \font\fiveeufm=eufm5
\font\teneusm=eusm10 \font\seveneusm=eusm7
\font\fiveeusm=eusm5 \or \font\teneufm=eufm10 scaled
\magstephalf \font\seveneufm=eufm7 \font\fiveeufm=eufm5
\font\teneusm=eusm10 scaled \magstephalf
\font\seveneusm=eusm7 \font\fiveeusm=eusm5 \or
\font\teneufm=eufm10 scaled \magstep1 \font\seveneufm=eufm7
\font\fiveeufm=eufm5 \font\teneusm=eusm10 scaled \magstep1
\font\seveneusm=eusm7 \font\fiveeusm=eusm5 \fi

\newfam\eufmfam \newfam\eusmfam \textfont\eufmfam=\teneufm
\scriptfont\eufmfam=\seveneufm
\scriptscriptfont\eufmfam=\fiveeufm
\textfont\eusmfam=\teneusm \scriptfont\eusmfam=\seveneusm
\scriptscriptfont\eusmfam=\fiveeusm

\def\frak{\ifmmode\let\next\frak@\else
 \def\next{\errmessage{Use \string\frak\space only in math
 mode}}\fi\next} \def\frak@#1{{\frak@@{#1}}}
 \def\frak@@#1{\fam\eufmfam#1} 
 \def\sh{\ifmmode\let\next\sh@\else
 \def\next{\errmessage{Use \string\sh\space only in math
 mode}}\fi\next} \def\sh@#1{{\sh@@{#1}}}
 \def\sh@@#1{\fam\eusmfam#1}

\ifcase\@ptsize \font\tenmsa=msam10 \font\sevenmsa=msam7
 \font\fivemsa=msam5 \font\tenmsb=msbm10
 \font\sevenmsb=msbm7 \font\fivemsb=msbm5 \or
 \font\tenmsa=msam10 scaled \magstephalf
 \font\sevenmsa=msam7 \font\fivemsa=msam5
 \font\tenmsb=msbm10 scaled \magstephalf
 \font\sevenmsb=msbm7 \font\fivemsb=msbm5 \or
 \font\tenmsa=msam10 scaled \magstep1 \font\sevenmsa=msam7
 \font\fivemsa=msam5 \font\tenmsb=msbm10 scaled \magstep1
 \font\sevenmsb=msbm7 \font\fivemsb=msbm5 \fi

\newfam\msafam \newfam\msbfam \textfont\msafam=\tenmsa
\scriptfont\msafam=\sevenmsa
\scriptscriptfont\msafam=\fivemsa \textfont\msbfam=\tenmsb
\scriptfont\msbfam=\sevenmsb
\scriptscriptfont\msbfam=\fivemsb

\def\Bbb{\ifmmode\let\next\Bbb@\else
 \def\next{\errmessage{Use \string\Bbb\space only in math
 mode}}\fi\next} \def\Bbb@#1{{\Bbb@@{#1}}}
 \def\Bbb@@#1{\fam\msbfam#1} \def\hexnumber@#1{\ifnum#1<10
 \number#1\else \ifnum#1=10 A\else\ifnum#1=11
 B\else\ifnum#1=12 C\else \ifnum#1=13 D\else\ifnum#1=14
 E\else\ifnum#1=15 F\fi\fi\fi\fi\fi\fi\fi}
 \def\msa@{\hexnumber@\msafam} \def\msb@{\hexnumber@\msbfam}
 \mathchardef\square="0\msa@03

\makeatother

\newcommand{\ZZ}{{\Bbb Z}} 
 
\newcommand{\EE}{{\Bbb E}}
\begin{document}
 
\title{
Geometric Discretisation of the Toda System}

\author{ Adam Doliwa\\
Istituto Nazionale di Fisica Nucleare, Sezione di Roma\\
P-le Aldo Moro 2, I--00185 Roma, Italia\\ 
and\\ 
Institute of Theoretical Physics, Warsaw University\\
ul. Ho\.{z}a 69, 00-681 Warsaw, Poland \thanks{Permanent address}}

\date{}
\maketitle

\begin{abstract}
\nin The Laplace sequence of the discrete conjugate nets is 
constructed. The invariants of the nets satisfy, in full analogy to the 
continuous case, the system of difference equations equivalent to the 
discrete version of the generalized Toda equation.

\bigskip

\nin {\it Keywords:} Integrable systems, discrete geometry, Toda equation
\end{abstract}


\vskip1cm

Preprint ROME1-1160/96, October 27, 1996

Dipartimento di Fisica, Universit\`a di Roma "La Sapienza"

I.N.F.N. -- Sezione di Roma

\vfill

\pagebreak
\section{Introduction}
Among all the integrable lattice systems a particular role is played by
the discrete analogue of the generalized Toda system
\cite{LPS}\cite{Hir}\cite{DJM}. Its suitable limits with respect to the 
lattice parameters give rise to the various types of soliton equations
including the Kadomtsev--Petviashvili equation, Korteweg--de Vries
equation, Benjamin--Ono equation, sine-Gordon equation or their
differential-difference versions.
Moreover, the discrete Toda system provides unification of
many particular examples of solvable models of statistical physics
and quantum field theory \cite{KNS}\cite{KLWZ}\cite{Kor}.

To construct integrable discretisation of the given soliton equation
there have been used several different approaches, for example:
\begin{itemize}
\item discrete version of the Lax system \cite{AbLad1}
\item Hirota's method via the bilinear form \cite{Hir}
\item extension of the Zakharov -- Shabat dressing method \cite{LPS}
\item direct linearisation using linear integral equations \cite{NQC1}.
\end{itemize}
Recently another discretisation approach has been applied to those of
soliton equations which describe geometrically meaningful objects such as
curves or some special types of surfaces \cite{DS1}--\cite{BP2}. As a working
principle, the discrete analogues of the relevant geometric properties
must be found.

This approach has been used to discretise Toda system. In section 2,
we collect useful facts from the theory of the conjugates nets (CNs) \cite{Eis}
and recall their connection with the Toda system \cite{Weiss}. 
In section 3, we
construct the Laplace sequence of the discrete conjugate nets (DCNs). In
the section 4, we define invariants of the sequence and write down
a difference equation relating the invariants of the neighboring DCNs
which is the discrete version of the Toda system.

\section{Conjugate nets, their Laplace transform 
and Toda system}
{\bf Definition 1.} Two directions tangent to the surface $M\subset
{\EE}^3$ in point $\xx\in M$ are called conjugate when they are conjugate
with respect to the Dupin indicatrix for the point \cite{Eis}.

\bigskip

\nin {\bf Fact 1.} Given a curve $\gamma$ on a surface $M\subset {\EE}^3$. The 
ruled surface made out of lines conjugate to tangent directions to
$\gamma$ is developable, i.e. it is formed by the tangent directions to
a curve $\gamma_{1}$.

\bigskip

\epsffile{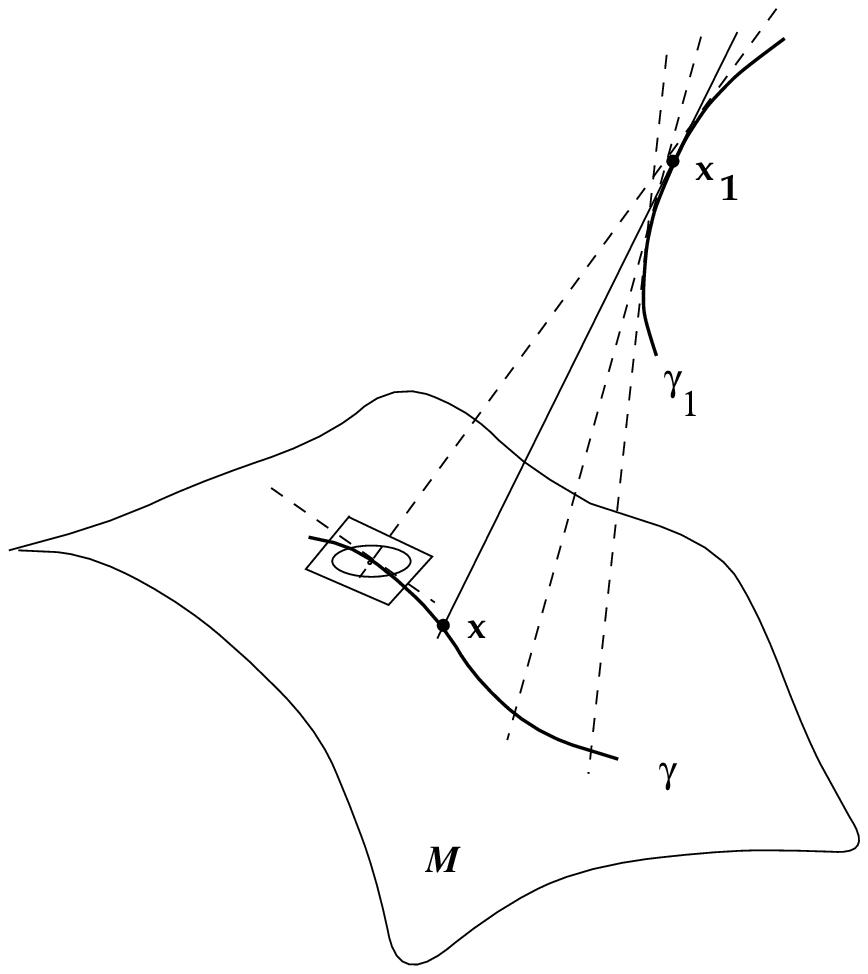}

\bigskip

{\bf Fig. 1}

\bigskip

\nin {\bf Definition 2.} Two families of curves on a surface $M\subset {\EE}^3$
are said to form conjugate net if the tangents to a curve of each family
at their point of intersection have conjugate directions.

\bigskip

\nin The curves of the first family due to the Fact 1. give rise to a
surface $M_{1}$. Similarly, the curves of the second family give rise to
a surface $M_{-1}$.

\bigskip

\nin {\bf Fact 2} The points of both new surfaces which correspond to
initial families of curves on $M$ form again conjugate nets.

\bigskip

\nin Similarly, $M_{1}$ gives rise to a new conjugate net $M_{2}$ (one can show
that $M_{(1)-1} = M$). This way we obtain (in general infinite) sequence
of conjugate nets
\[ ..., M_{-2},\;  M_{-1},\;  M_{0} = M, \; M_{1},\;  M_{2}, ... \]
called the Laplace sequence.
The analytic expressions for these results are based on the following

\bigskip

\nin {\bf Fact 3.} A necessary and sufficient condition for the
parametric curves of coordinates $(u,v)$ to form a conjugate system is that
the second mixed derivative of the vector $\xx(u,v)$ pointing the surface is
tangent to the surface.

\bigskip

\nin This means that the vector-function $\xx(u,v)$ satisfies the
Laplace equation
\beq \xx_{,uv} + a \xx_{,u} + b \xx_{,v} = 0 \; \; \; ,
\label{eq:Lapl} \eeq
where $a$ and $b$ are determinate functions of the parameters $u,v$ , and comma 
denotes differentiation.

To construct the surface $M_{1}$ we use the lines of constant $v$ on
$M$. Then the points of $M_{1}$ are given by
\beq \xx_{\bf 1}= \xx + \lambda \xx_{,u} \; \; ,
\label{eq:x1poi} \eeq
where $\lambda$ is a determinate function of the parameters $u,v$. But
the direction from $\xx$ to $\xx_{\bf 1}$ must be tangent to the line
$u=\const$ on $M_{1}$
\beq \xx_{{\bf 1},v} = \mu \xx_{,u} \; \; ,
\label{eq:x1tan} \eeq
where again $\mu$ is a function of $u,v$.

As a consequence of the equations (\ref{eq:Lapl}), (\ref{eq:x1poi}) and
(\ref{eq:x1tan}) we obtain an overdetermined system
\beq (\lambda_{,v} + a\lambda - \mu)\xx_{,u} + (1-b\lambda)\xx_{,v} = 0 \; \; ,
\label{eq:overdet} \eeq
which implies
\beq \lambda=\frac{1}{b} \; \; \; , \; \; \; \mu = \left(\frac{1}{b}
\right)_{,v} - \frac{a}{b} \; \; .
\eeq
Finally
\bea \xx_{\bf 1} & = & \xx + \frac{1}{b} \xx_{,u} \nonumber \; \; ,\\
\xx_{{\bf 1},v} & =  & \left(  \left(\frac{1}{b}
\right)_{,v} - \frac{a}{b}\right) \xx_{,u} \; \; ,
\label{eq:x1}\eea
and similarly 
\bea \xx_{\bf -1} & = & \xx + \frac{1}{a} \xx_{,v} \; \; , \nonumber \\
\xx_{{\bf -1},u} & =  & \left(  \left(\frac{1}{a}
\right)_{,u} - \frac{b}{a}\right) \xx_{,v} \; \; .
\label{eq:x-1} \eea
To proceed further we are forced to leave the domain of the theory of
surfaces in the Euclidean space $\EE^3$. It turns out that the notion of
the conjugate net is a projective property, i.e. when the whole space is
subjected to the projective transformation CNs are
transformed into CNs. As the Cartesian coordinates of the
points of the surface and functions $a,b$ change, it is important to
know the projective invariants of the conjugate nets. Such invariant
functions are
\bea  k & = & b_{,v} + ab \; \; \; , \nonumber \\
\label{eq:kh} h & = & a_{,u} + ab \; \; \; .
\eea 
Let us find the relation between the invariants of the Laplace sequence
of conjugate nets. It is easy to show, using equations (\ref{eq:x1}),
that the vector-function $\xx_{\bf 1}$ satisfies the Laplace equation
again (proving the Fact 2)
\beq \xx_{{\bf 1},uv} + \frac{k}{b} \xx_{{\bf 1},u} + \left( b - 
\left(\ln\frac{k}{b} \right)_{,u}\right) \xx_{{\bf 1},v} = 0 \; \; \; ,
\label{eq:Lapl1} \eeq					
which implies the form of the transformed coefficients $a,b$
\bea a_{1} & = & \frac{k}{b} \nonumber \;  \; \\
\label{eq:ab1} b_{1} & = & b - \left(\ln\frac{k}{b} \right)_{,u} \; \; .
\eea
In consequence, the Laplace transform of the invariants reads
\bea \label{eq:h1} h_{1} & = & k \; \; \; , \\
\label{eq:k1} k_{1} & = & 2 k  - (\ln k)_{,uv} - h \; \; .
\eea
Finally, the sequence $k_{l}$ of invariants satisfies the system of
equations
\beq 
(\ln k)_{,uv} = - k_{l+1} + 2 k_{l} - k_{l-1} \; \; \; , \; \; \; 
l = ... -2, \; -1, \; 0, \; 1, \; 2, ...
\label{eq:TodaDarb}
\eeq
known in this context to Darboux \cite{Darb}, and equivalent to the standard 
form of the
Toda system
\beq \theta_{l,uv} = - \ee^{\theta_{l+1} - \theta_{l}} +
\ee^{\theta_{l} - \theta_{l-1}} \; \; ,
\label{eq:Toda} \eeq
where
\beq k_{l} = \ee^{\theta_{l+1} - \theta_{l}} \; \; .
\eeq
The next two sections are devoted to the discretisation of all the reasoning
presented above.

\section{Discrete conjugate nets and their Laplace sequences}
{\bf Definition 3.} By a discrete conjugate net we mean a map $\XX: \ZZ^{2} \ra 
\EE^3$ such that all the elementary quadrilaterals
\[ \langle \XX(i,j),\XX(i+1,j),\XX(i,j+1),\XX(i+1,j+1) \rangle \] 
are planar (see
\cite{Sauer}).

\bigskip

\nin The above definition is a natural discretisation of the basic
property of CNs described in the Fact 3. The discrete analogue of the
Laplace equation (\ref{eq:Lapl}) is
\bea \XX(i+1,j+1) &  = & \XX(i,j) + A(i,j)(\XX(i+1,j) - \XX(i,j)) \nonumber \\
 &  + &  B(i,j)(\XX(i,j+1) - \XX(i,j))  \; \; \; .
\label{eq:DLapl} \eea
From now on, for any function $F$ on $\ZZ^2$ we will use the following
convention
\beq F^q_p = F(i+p,j+q) \; \; ,
\eeq
for example, the above equation (\ref{eq:DLapl}) can be rewritten as
\[ \XX^1_1 = \XX + A(\XX_1 - \XX) + B(\XX^1 - \XX) \; \; . \]
The intersection points of the directions of the opposite edges of the
elementary quadrilateral (they may intersect at the plane in 
infinity --- we must take
into consideration the projective picture) define points of the Laplace
transforms of the net, denoted by $\tXX$ and $\cXX$ (see Fig. 2 where also 
the convention of labelling points of the new nets is given).

\bigskip

\epsffile{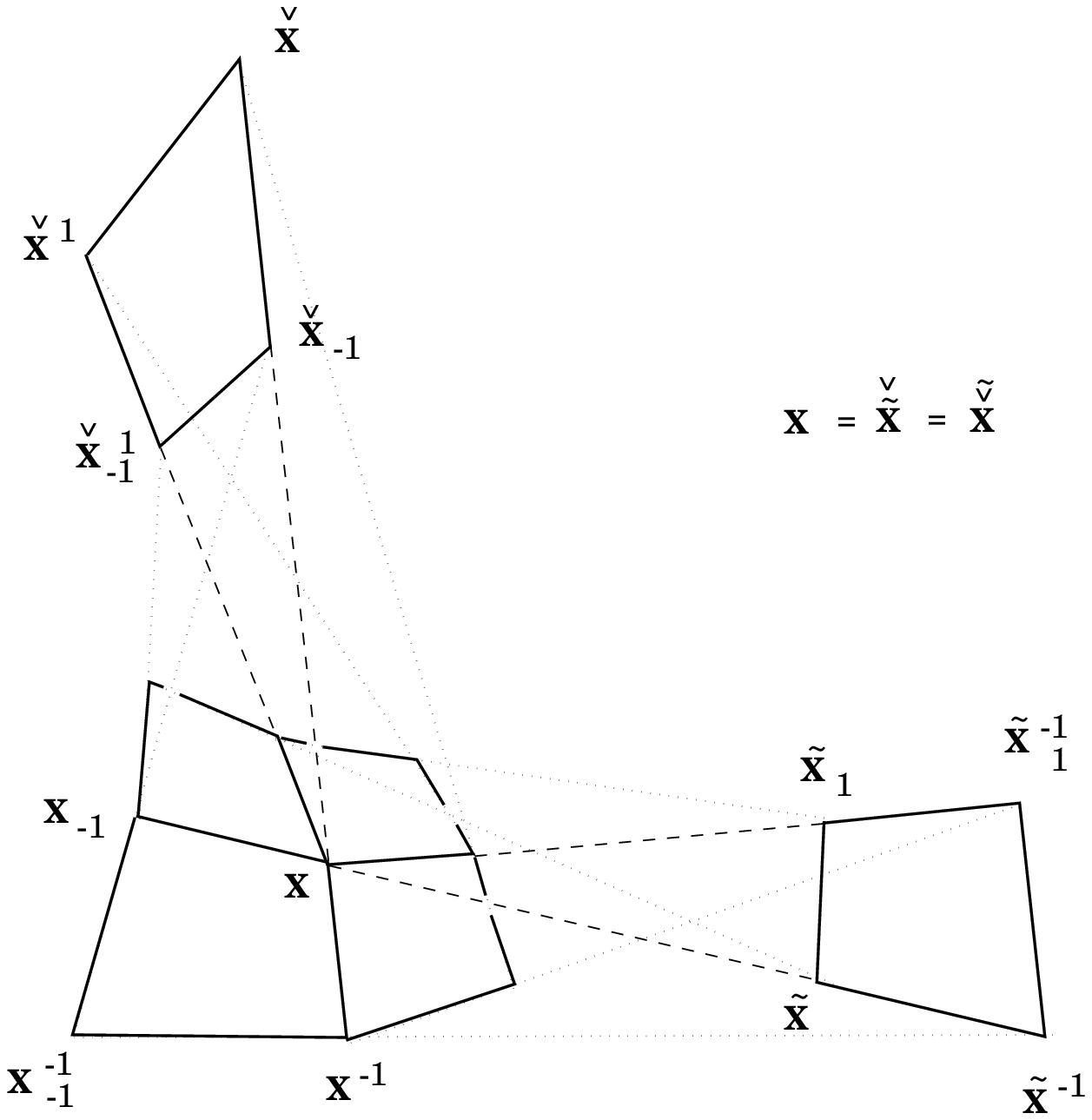}

\bigskip

{\bf Fig. 2}

\bigskip

\nin The following two facts can be easily examinated.

\bigskip

\nin {\bf Fact 4.} Both Laplace transforms $\tXX$ and $\cXX$ of the
discrete conjugate net $\XX$ are again discrete conjugate nets.

\bigskip

\nin {\bf Fact 5.} The Laplace transformations of the discrete conjugate
nets do not branch, i.e. $\check{\tXX} = \tilde{\cXX} = \XX$.

\bigskip

\nin This way any DCN gives rise to a whole sequence (in general
infinite on both sides) of DCNs.

In what follows, we define the analogues of the functions $\lambda, \mu$
and find the Laplace transforms of the coefficients $A, B$.
\bea \tXX & = & \XX_{-1} + \Lambda(\XX - \XX_{-1}) \; \; , \nonumber \\
\tXX - \tXX^{-1} & = & M (\XX - \XX_{-1}) \; \; .
\label {eq:DtX} \eea
Similarly as for CNs we can express functions $\Lambda, M$ in terms of
$A$ and $B$
\bea \Lambda_1 & = & \frac{A}{1-B} \; \; \; , \nonumber \\
M^1_1 & = &   \frac{A^1}{1-B^1} -  \frac{1}{1-B} \; \; .
\label{eq:LaM} \eea
Let us discuss, for a moment, the limit of the small lattice parameter
$\eps$ (in general one can take two independent lattice parameters) 
\bea \XX_1 - \XX & \simeq & \eps\XX_{,u} \; \; \; , \nonumber \\
\XX^1 - \XX & \simeq & \eps\XX_{,v} \; \; \; , \\
\XX^1_1 - \XX^1 - (\XX_1 - \XX) & \simeq & \eps^2\XX_{,uv} \; \; \; .
\nonumber \label{eq:lim1} \eea
In consequence 
\bea 1- A & \simeq & \eps a \; \; \; , \nonumber \\
 1- B & \simeq & \eps b \; \; \; , \nonumber \\
\Lambda & \simeq & \frac{\lambda}{\eps} \; \; , \\
M & \simeq & \mu \; \; \; \; . \nonumber
\label{eq:lim2} \eea
The equations (\ref{eq:DtX}) can be used to obtain the discrete Laplace
equation for the net $\tXX$ (proving the Fact 4) and allow to find the
transformed coefficients $\tilde{A}$ and $\tilde{B}$
\bea \tilde{A} & = & \frac{\Lambda^1_1}{\Lambda^1_1 - M^1_1} \; \; ,
\nonumber \\
\tilde{B} & = & \frac{M^1_1}{M^1}\cdot 
\frac{M^1 -\Lambda^1 + 1}{M^1_1 - \Lambda^1_1} \; \; .
\label{eq:tAB} \eea
Similar considerations can be made also for the second Laplace transform
$\cXX$ of the conjugate net $\XX$.

\section{Invariants of the discrete conjugate nets and the discrete Toda
system}
The fundamental property of the projective transformations is that straight 
lines are transformed into straight lines and consequently planes into
planes. This implies that the notion of discrete conjugate net is in
fact projective notion, and Laplace sequences are mapped into
Laplace sequences od DCNs.

In looking for the discrete analogues of the invariants $k$ and $h$ let us
recall that the basic projective invariant is the cross-ratio of 
four points $A,B,C,D$ on the line
\beq Q(A,B,C,D) = \frac{\vec{AB}}{\vec{BC}} \div \frac{\vec{AD}}{\vec{CD}}
\; \; ,
\eeq
where the ratio of two vectors should be understood as a proportionality factor.  

\bigskip

\nin {\bf Remark 1.} 
\beq Q(A,B,C,D) = Q(D,C,B,A) \; \; \; . \eeq

\bigskip
\nin In our construction of the discrete conjugate nets and their Laplace 
sequences there are two natural systems of four points on the line 
(see Fig. 2), and we propose for the invariants
\bea K & = & Q(\XX_{-1},\XX,\tXX,\tXX^{-1}) \; \; , \nonumber \\
H & = & Q(\XX^{-1},\XX,\cXX,\cXX_{-1}) \; \; .
\label{eq:KH} \eea
Equations (\ref{eq:DtX}) and (\ref{eq:LaM}) allow to express the invariant $K$ 
in terms of the coefficients $A,B$ of the Laplace equation
\beq
\label{eq:KAB}
K = \frac{M}{(\Lambda -1)(M-\Lambda)} = 
\frac{A_{-1}B^{-1}_{-1}}{A_{-1}+B_{-1}-1 } -1 \; \; , \eeq
and similarly
\beq
\label{eq:HAB}
H = \frac{B^{-1}A^{-1}_{-1}}{A^{-1}+B^{-1}-1 } -1 \; \; . \eeq
In the limit of the small lattice parameter we obtain
\bea K & \simeq & \eps^2 k \; \; \; , \nonumber \\
H & \simeq & \eps^2 h \; \; \; .
\label{eq:lim3} \eea
To construct the discrete Toda system we need to find the Laplace
transforms of the invariants. Using formulas
(\ref{eq:LaM}),(\ref{eq:tAB}),(\ref{eq:KAB}) and (\ref{eq:HAB}) we derive the
following relations
\beq \tilde{H} = K \;\; \; , \eeq
\beq \tilde{K_1}+1 = \frac{(K^1_1 + 1)(K + 1)}{H^1 + 1}
\frac{K_1 K^1}{K^1_1 K} \; \; . \label{eq:tK} \eeq
The first of the above equations is just the consequence of the
Remark 1. Finally, the equation (\ref{eq:tK})
rewritten in terms of functions $K(i,j,l)$ (we go back to the full definitions)
\beq ..., \; K(i,j,-1) =\check{K}, \;  K(i,j,0)=K, \; K(i,j,1) =
\tilde{K}, \; .... \eeq
gives rise to the difference system
\beq \frac{(K(i+1,j,l+1) + 1)(K(i,j+1,l-1) + 1)}
{(K(i+1,j+1,l) + 1)\:(K(i,j,l) + 1)} = 
\frac{K(i+1,j,l)K(i,j+1,l)}{K(i+1,j+1,l)K(i,j,l)} \; \; . \eeq
The above equations in the limit of the small parameter $\eps$ reduce to
the equations (\ref{eq:TodaDarb}). They are equivalent to the known in the
literature gauge invariant discrete version of the Toda system 
(see e.g. \cite{KLWZ}).

\bigskip

\nin{\bf Remark 2.} In all the above calculations we had not used essentially 
the fact that $\XX$ is a point of the space of dimension $3$. All the results 
remain valid if we consider the ambient space of the bigger dimension.

\bigskip

\nin{\bf Remark 3.} The transformation of a given discrete Laplace 
equation to a new such equation can be considered also as a discrete
analog of the {\it cascade method} (proposed by Laplace) of solving
the equation. This method is especially useful when the Laplace sequence
terminates on a surface degenerated to a curve. 

\bigskip

\nin{\bf Remark 4.} The finite Laplace sequences correspond to the open
Toda systems, which are related to the $A$-type 
simple Lie algebras, periodic Laplace sequences are related in turn to the 
affine $A^{(1)}$ Lie algebras.
Various additional geometric structures
in the ambient space allow to find the open and periodic Toda systems
related to other simple Lie algebras (see e.g. \cite{AD1},\cite{AD2}).
This procedure can be applied also to the discrete Toda system and should
recover equations existing in the literature \cite{KNS}\cite{Ward}.

\section{Acknowledgments}
I would like to thank A. Sym for pointing out reference \cite{Sauer} and
to P. M. Santini for stimulating discussions.
This work was supported partially by the Committee of Scientific Research (KBN)
under the Grant Number 2P03 B 18509.

\end{document}